# Study on layout of double rotated serpentine springs for vertical-comb-driven torsional micromirror

**Biyun Ling[1], Yuhu Xia[1,2], Minli Cai[1,2], Xiaoyue Wang[1] and Yaming Wu[1,2]**

[1] State Key Laboratory of Transducer Technology, Shanghai Institute of Microsystem and Information Technology, Chinese Academy of Sciences, Shanghai 200050, China
[2] University of Chinese Academy of Sciences, Beijing 100049, China

E-mail: lingbiyun@mail.sim.ac.cn

**Abstract.** The combination of double rotated serpentine springs (RSS) and vertical comb-drive is a suitbale solution for the development of torsional micromirror with high fill factor, low fabrication difficulty and good performance. However, the alignment error between upper and lower comb set caused by fabrication can induce force with unexpected direction. And the cross-axis coupled spring constants in double rotated serpentine springs (DRSSs) makes micromirror more susceptible to this alignment error. Herein, in order to minimize the unexpected deflection caused by alignment error of vertical-comb-driven micromirror, this paper, for the first time, studies the effect of layout (centrosymmetrically-arranged and axisymmetrically-arranged) of DRSSs on cross-axis coupled spring constants. Both of theoretical analysis and finite element analysis (FEA) simulation are conducted to reveal this phenomenon. With an example, centrosymmetrically-arranged DRSSs are proved to be more resistant to pull-in of two comb sets. Finally, the relationship between key structure parameters and cross-axis coupled spring constants of centrosymmetrically -arranged DRSSs are presented.

## 1. Introduction

MMA, as its name implies, consists of periodically-arranged micromirrors that can be actuated to control the direction and the phase of light. It is always deemed as the core part in optic-based systems that need simultaneous manipulation of multiple light beams, and is widely used in many industrial and scientific fields [1].

Fill factor, defined as the ratio of reflector area to chip area of each micromirror unit, is one of the major concerns in developing MMA. To be specific, high fill factor means low optical loss and low possibility of stray light for MMA, both of which can reduce the signal to noise ratio of reflected light beam, and thereby greatly improve the performances of optical systems that are equipped with MMA.

These optical systems, of course, can employ extra optical elements (such as collecting lens and light amplifier), rather than utilizing MMA, to achieve the same performances, while their complexity, size and cost increases inevitably.

To maximize fill factor, reflector of each micromirror should be arranged on the topmost tier, and supportive structures (including spring and gimbal) should be hidden below reflector. However, it increases the number of structures layers of MMA, and correspondingly increases the difficulty and failure risk of MMA fabrication (either bulk silicon process, or surface process, or micro-assembly). Thus, comparatively, optimizing the shape and layout of supportive structures so as to make supportive structures and reflector be arranged on the same structure layer seems to be a more realistic solution. For such case, serpentine spring, getting its name from the meandering snake-like pattern of the spring segment, is well suited, since it allows small spring constant with a reasonable occupation of area [2]-[5]. Besides, the small spring constant of serpentine spring is beneficial to reduce the required driving strength of actuator of micromirror, which can make electrostatic actuation more available (compared to other actuation methods, electrostatic actuation is more stable, more repeatable and more compatible to MEMS processing).

Herein, aimed at developing MMA with high fill factor, low fabrication difficulty and good performance, this paper studies single-axis torsional micromirror with DRSSs and vertical comb-drive, as shown in Figure 1. To the best of my knowledge, this structure has two potential risks. Firstly, although vertical comb-drive can achieve lower driving voltage and larger stroke than plate drive, the alignment error between upper comb set and lower comb set caused by fabrication can induce force with unexpected direction, which may lead to pull-in of these two comb sets. Secondly, there exists cross-axis coupled spring constants in DRSSs, which makes this micromirror more susceptible to the force generated from alignment error of vertical comb-drive.

Therefore, in order to minimize the unexpected deflection caused by alignment error of vertical-comb-driven micromirror, this paper, for the first time, studies the effect of layout (centrosymmetrically-arranged and axisymmetrically-arranged) of DRSSs on cross-axis coupled spring constants.

## 2. Structure Design and Fabrication

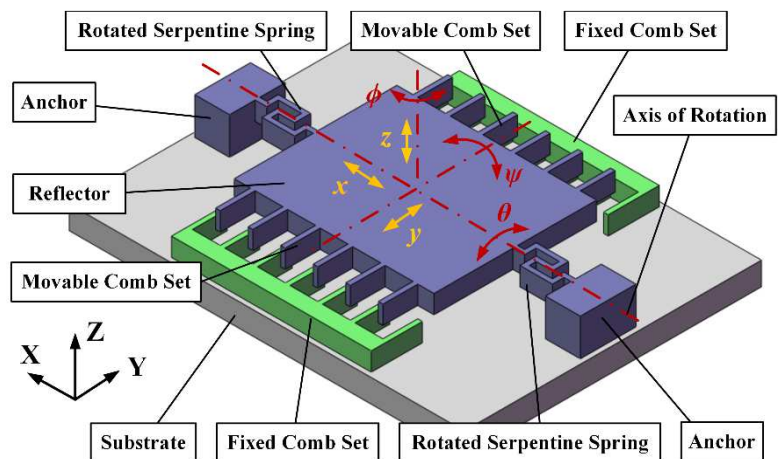

Figure 1: Schematic diagram of torsional single-axis micromirror with vertical comb-drive and DRSSs (DRSSs are axisymmetrically-arranged).

Figure 1 shows the schematic diagram of proposed torsional single-axis micromirror. With suspension of DRSSs, the reflector is connected to the anchors on both sides. Two movable comb sets, reflector and RSSs are arranged on the upper layer, while two fixed comb sets are arranged on the lower layer. Movable comb sets along with reflector are grounded electrically. When a voltage is applied to a fixed comb set, electrostatic force is generated between this fixed comb set and its corresponding movable comb set, which drives the reflector rotate along the rotation axis ($\theta$).

The fabrication process of this micromirror is illustrated in Figure 2. A SOI (silicon-on-insulator) wafer and a silicon wafer are adopted in this process. (a) Au pads are deposited and patterned on the device layer of SOI wafer. (b) Combs are patterned and etched on the device layer of SOI wafer. (c) Thermal oxidization is conducted to form $SiO_2$ film on the silicon wafer. (d) Au pads are deposited and patterned on one side of silicon wafer. (e) The SOI wafer is bonded to the silicon wafer with processing of Au-Au bonding. (f) The substrate layer and buried $SiO_2$ layer of SOI wafer are removed. (g) The device layer of SOI wafer is patterned and etched, forming movable combs, RSSs and reflector. It should be mentioned that there inevitably exists alignment error in step (g), regardless of alignment method (either front-side alignment and back-side alignment). Thus, each movable comb does not locate right in the middle of its adjacent two fixed combs. In this case, force along X-axis can be generated, as shown in Figure 1, which leads to unexpected deflection.

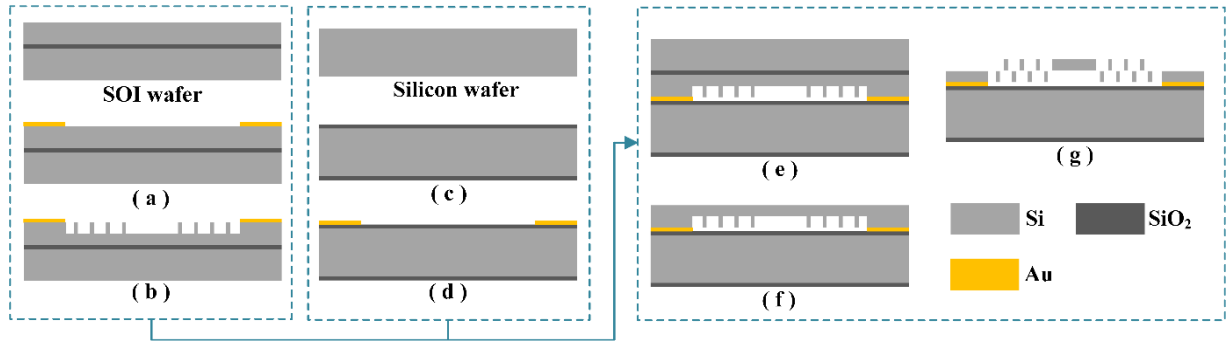

Figure 2: Main steps of fabrication process.

## 3. Theoretical Modeling

*3.1 Spring Constants of RSS*

The top view of RSS is shown in Figure 3. Herein, the alignment-error-induced force is along X-axis, so this paper studies the spring constants related with in-plane movement ($\delta_x$, $\delta_y$ and $\delta_\varphi$). The relationship between in-plane movement and applied force ($F_x$ and $F_y$) or moment ($M_\varphi$) can be expressed as:

$$\begin{bmatrix} F_x \\ F_y \\ M_\varphi \end{bmatrix} = \begin{bmatrix} k_x & k_{xy} & k_{x\varphi} \\ k_{xy} & k_y & k_{y\varphi} \\ k_{x\varphi} & k_{y\varphi} & k_\varphi \end{bmatrix} \begin{bmatrix} \delta_x \\ \delta_y \\ \delta_\varphi \end{bmatrix} \quad (1).$$

There are 6 independent terms, since this matrix is symmetric. The elements of this spring constants matrix are calculated by using the principle of virtual work with the unit-load method [6]. Notably, while using unit-load method to calculate this matrix, some assumptions are made, including small-angle approximation, only bending moment, and ignorance of deflection from shear, spring shortening and

lengthening. Thus, the total energy of RSS is expressed as:

$$U = \sum \int_l \frac{M(\xi)^2}{2EI} d\xi,$$

where $l$ is the length of each spring segment, $\xi$ is the infinitesimal along each spring segment. $E$ is Young's modulus and $I$ is moment of inertia.

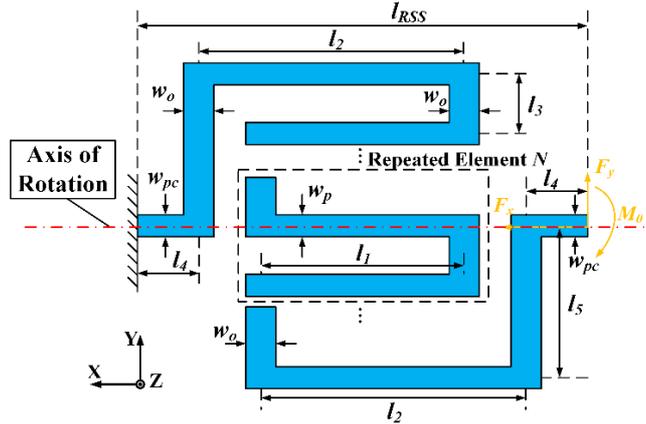

Figure 3: Top view of RSS. The labels show the design variables.

*3.2 Spring Constants of DRSSs*

Plotted in Figure 4 is the top view of centrosymmetrically-arranged and axisymmetrically-arranged DRSSs. The former one consists of Type A and Type B RSS, while the latter one merely consists of two Type A RSSs. It should be mentioned that Type A and Type B RSS are symmetric to each other along X-axis. After calculating the spring constants of Type A and Type B RSS, we find that all their spring constants are identical, except $k_{xy}$ and $k_{x\varphi}$ ($k_{xy-\text{Type A}}=-k_{xy-\text{Type B}}$ and $k_{x\varphi-\text{Type A}}=-k_{x\varphi-\text{Type B}}$). Thus, the spring constants of DRSSs can be achieved, as listed in Table 1.

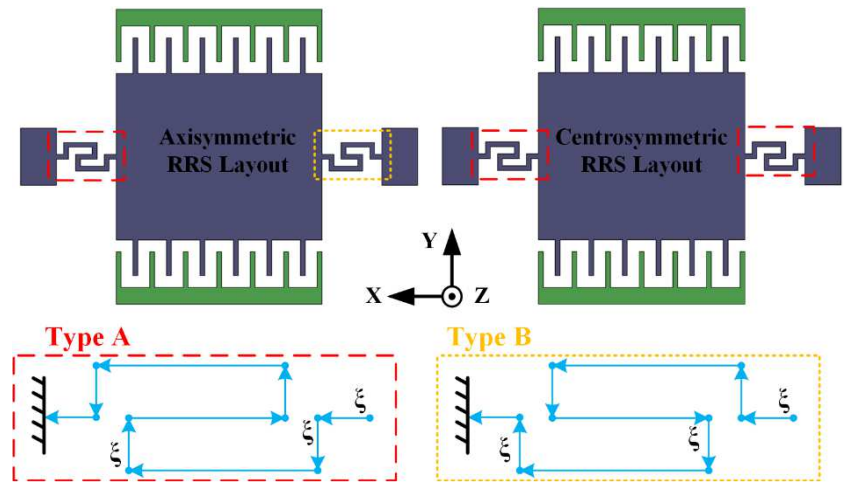

Figure 4: Schematic diagram of axisymmetrically-arranged and centrosymmetrially-arranged DRSSs. Arrows in the inserted figures represent the integral direction for each spring segment while using energy-based Castiliano's method to calculate the total energy of the spring.

Table 1: Summary of Spring Constants of Centrosymmetrically-arranged and Axisymmetrically-arranged DRSSs.

| | Centrosymmetrically-arranged DRSS | Axisymmetrically-arranged DRSS |
|---|---|---|
| $K_x$ | $2k_x$ | $2k_x$ |
| $K_y$ | $2k_y$ | $2k_y$ |
| $K_\varphi$ [7] | $2k_\varphi + 6k_\varphi(1+l_m/l_{RSS})^2$ | $2k_\varphi + 6k_\varphi(1+l_m/l_{RSS})^2$ |
| $K_{xy}$ | $2k_{xy}$ | 0 |
| $K_{x\varphi}$ | 0 | $2k_{x\varphi} + 6k_{x\varphi}(1+l_m/l_{RSS})^2$ |
| $K_{y\varphi}$ | 0 | 0 |

It is obvious that the spring constants of centrosymmetrically-arranged and axisymmetrically-arranged DRSSs are the same, except $K_{xy}$ and $K_{x\varphi}$. To be specific, when a force along X-axis is applied to the micromirror with centrosymmetrically-arranged DRSSs, this micromirror will move along Y-axis. Similarly, when this force is applied to the micromirror with axisymmetrically-arranged DRSSs, this micromirror will rotate around Z-axis. Therefore, similar to equation (1), the relationship between in-plane movement of this micromirror ($\delta_{x\text{-DRSS}}$, $\delta_{y\text{-DRSS}}$ and $\delta_{\varphi\text{-DRSS}}$) and applied force ($F_x$ and $F_y$) or moment ($M_\varphi$) can be expressed as:

$$\begin{bmatrix} F_x \\ F_y \\ M_\varphi \end{bmatrix} = \begin{bmatrix} 2k_x & 2k_{xy} & 0 \\ 2k_{xy} & 2k_y & 0 \\ 0 & 0 & 2k_\varphi + 6k_\varphi(1+l_m/l_{RSS})^2 \end{bmatrix} \begin{bmatrix} \delta_{x\text{-DRSS}} \\ \delta_{y\text{-DRSS}} \\ \delta_{\varphi\text{-DRSS}} \end{bmatrix} \quad (2),$$

$$\begin{bmatrix} F_x \\ F_y \\ M_\varphi \end{bmatrix} = \begin{bmatrix} 2k_x & 0 & 2k_{x\varphi} + 6k_{x\varphi}(1+l_m/l_{RSS})^2 \\ 0 & 2k_y & 0 \\ 2k_{x\varphi} + 6k_{x\varphi}(1+l_m/l_{RSS})^2 & 0 & 2k_\varphi + 6k_\varphi(1+l_m/l_{RSS})^2 \end{bmatrix} \begin{bmatrix} \delta_{x\text{-DRSS}} \\ \delta_{y\text{-DRSS}} \\ \delta_{\varphi\text{-DRSS}} \end{bmatrix} \quad (3),$$

where the former is for micromirror with centrosymmetrically-arranged DRSSs and the latter is for micromirror with centrosymmetrically-arranged DRSSs.

## 4. Model Verification and Discussion

The model derived above is verified with FEA simulation, where vertical comb-drives are omitted in structure modeling. Force along X-axis ($F_x$) with strength ranging from 0 to 10μN is applied to the left side of micromirror, and cross-axis coupled Z-axis rotation angle ($\varphi$) and Y-axis displacement ($y$) of the micromirror are recorded, as shown in Figure 5. The simulated $\varphi$ and $y$ basically agree with calculated results, while the latter ones are about 20% smaller than the former ones. The differences between the calculated and simulated results could be attributed to merely considering bending moment of RSSs.

Notably, for a micromirror with high-aspect-ratio comb, in-plane rotation (Z-axis rotation) is more liable to pull-in of two comb sets [8]. Therefore, centrosymmetrically-arranged DRSSs are more suitable in this micromirror.

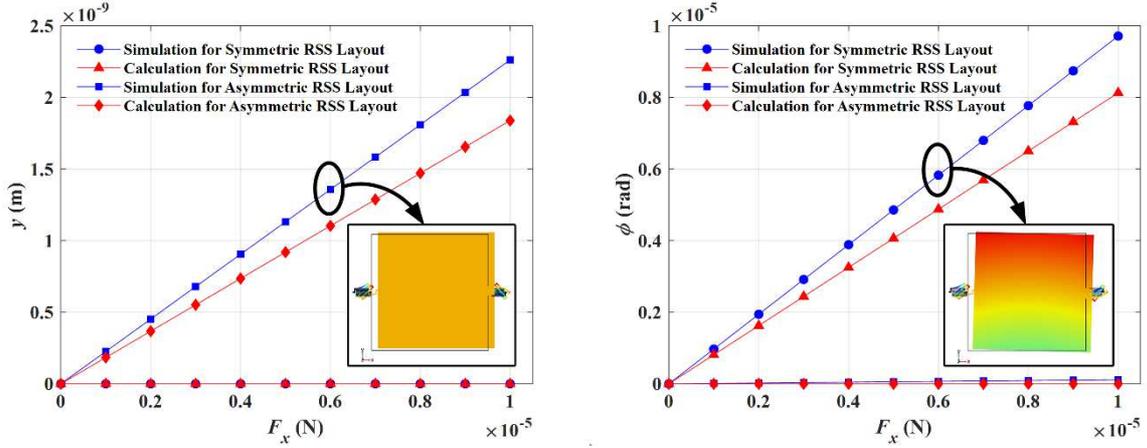

Figure 5: Simulation and calculation results of (a) $y$ and (b) $\varphi$ with respect to $F_x$ for micromirror with centrosymmetrically-arranged and axisymmetrically-arranged DRSSs. The dimension of micromirror is 400×400×30μm³. RSS structure parameters $N$, $l_1$, $l_2$, $l_3$, $l_4$, $l_5$, $w_o$, $w_p$ and $w_{pc}$ are 1, 40μm, 46μm, 6μm, 6μm, 12μm, 30μm, 3μm, 3μm and 3μm, respectively.

## 5. Parameter Optimization

The displacement along Y-axis ($\delta_{\text{y-DRSS-d}}$) caused by alignment-error-induced force ($F_x$) can be deduced from equation (2):

$$\delta_{\text{y-DRSS-d}} = -\frac{K_{xy}}{K_y}\frac{F_x}{K_x} = R_y \frac{F_x}{K_x}.$$

Herein, the absolute value of $R_y$ is minimized to reduce unexpected deflection. Thus, parameter sweep is conducted, and results are illustrated in Figure 6.

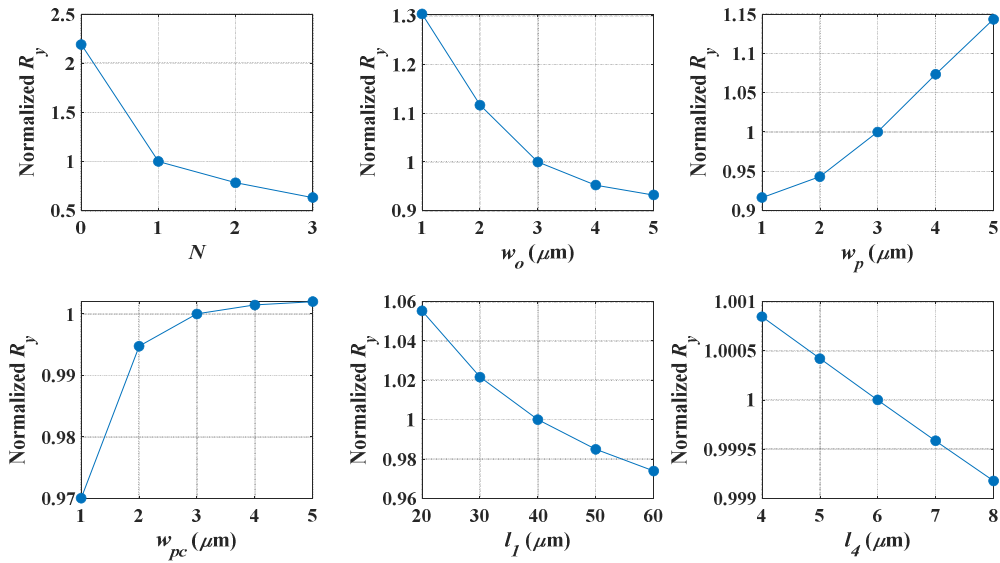

Figure 6: Relationship between $N$, $w_o$, $w_p$, $w_{pc}$, $l_1$, $l_4$ and normalized $R_y$. The dimension of micromirror is 400×400×30μm³. The initial RSS structure parameters $N$, $l_1$, $l_2$, $l_3$, $l_4$, $l_5$, $w_o$, $w_p$ and $w_{pc}$ are 1, 40μm, 46μm, 6μm, 6μm, 12μm, 30μm, 3μm, 3μm and 3 μm respectively. And initial $R_y$ is −0.24.

It is obvious that $|R_y|$ barely changes with $w_o$, $w_p$, $w_{pc}$, $l_1$ and $l_4$, and decreases with the increasing of $N$. Thus, when deciding the value of $w_o$, $w_p$, $w_{pc}$, $l_1$ and $l_4$, we may not take $R_y$ into account. Besides, although increasing $N$ can decrease $|R_y|$, we should take its influence to $K_x$ and $K_\theta$ into consideration.

**6. Conclusion**

this paper, for the first time, studies the effect of layout (centrosymmetrically-arranged and axisymmetrically-arranged) of DRSSs on cross-axis coupled spring constants. Both of theoretical analysis and finite element analysis (FEA) simulation are conducted to reveal this phenomenon. With an example, centrosymmetrically-arranged DRSSs are proved to be more resistant to pull-in of two comb sets. Finally, the relationship between key structure parameters and cross-axis coupled spring constants of centrosymmetrically-arranged DRSSs are presented. This study is beneficial to guide the design of micromirror with high fill factor, low fabrication difficulty and good performance, and thus can propel the development of MMA.

**Reference**

[1] Song, Y., Panas, R. M., and Hopkins, J. B. A review of micromirror arrays. *Precision Engineering*. 2018; 51: 729-761.

[2] Tang, W. C., Nguyen, T. C. H., and Howe, R. T. Laterally driven polysilicon resonant microstructures. *Sensors and actuators*. 1989; 20(1-2): 25-32.

[3] Peroulis, D., Pacheco, S. P., Sarabandi, K., and Katehi, L. P. Electromechanical considerations in developing low-voltage RF MEMS switches. *IEEE Transactions on microwave theory and techniques*. 2003; 51(1): 259-270.

[4] Cai, H., Ding, G., Yang, Z., Su, Z., Zhou, J., and Wang, H. Design, simulation and fabrication of a novel contact-enhanced MEMS inertial switch with a movable contact point. *Journal of Micromechanics and Microengineering*. 2008; 18(11): 115033.

[5] Sharma, A. K., and Gupta, N. Investigation of actuation voltage for non-uniform serpentine flexure design of RF-MEMS switch. *Microsystem technologies*. 2014; 20: 413-418.

[6] Barillaro, G., Molfese, A., Nannini, A., and Pieri, F. Analysis, simulation and relative performances of two kinds of serpentine springs. *Journal of Micromechanics and Microengineering*. 2005; 15(4): 736.

[7] Urey, H., Kan, C., and Davis, W. O. Vibration mode frequency formulae for micromechanical scanners. *Journal of Micromechanics and Microengineering*. 2005; 15(9): 1713.

[8] Lee, D., and Solgaard, O. Pull-in analysis of torsional scanners actuated by electrostatic vertical combdrives. *Journal of microelectromechanical systems*. 2008; 17(5): 1228-1238.